\documentclass[prl,aps,showpacs,superscriptaddress,twocolumn]{revtex4}
\usepackage{color} 
\usepackage{bm}
\usepackage{xcolor}
\usepackage{dcolumn}
\usepackage{times}
\usepackage{amssymb} 
\usepackage{amsmath} 
\usepackage{ragged2e}
\usepackage{graphicx}
\usepackage{epstopdf}
\usepackage[english]{babel}
\usepackage{mathrsfs}
\usepackage{textcomp}
\usepackage{amsthm}
\usepackage{amsfonts}
\usepackage{soul}
\usepackage{braket} 
\usepackage{hyperref}
\usepackage{cleveref}
\usepackage{float}
\newcommand{\be}{\begin{equation}}
\newcommand{\ee}{\end{equation}}
\newcommand{\bea}{\begin{eqnarray}}
\newcommand{\eea}{\end{eqnarray}}
\newcommand{\ba}{\begin{array}}
\newcommand{\ea}{\end{array}}

\usepackage{tikz}
\usetikzlibrary{patterns}
\usepackage[caption=false]{subfig}
\begin{document}

\title{Homogeneous and domain wall topological Haldane conductors with dressed Rydberg atoms}

\author{Arianna Montorsi}
\affiliation{Institute for condensed matter physics and 
complex systems, DISAT, Politecnico di Torino, I-10129, Italy}
\author{Serena Fazzini}
\affiliation{Physics Department and Research Center OPTIMAS, University of Kaiserslautern, D-67663 Kaiserslautern, Germany}
\author{Luca Barbiero}
\affiliation{Center for Nonlinear Phenomena and Complex Systems,
Universit\'e Libre de Bruxelles, CP 231, Campus Plaine, B-1050 Brussels, Belgium}

\date{\today}

\begin{abstract}
The interplay between antiferromagnetic interaction and hole motion is capable of inducing conducting Haldane phases with topological features described by a finite non-local string order parameter. Here we show that these states of matter are captured by the one dimensional $t-J_z$ model which can be experimentally realized with dressed Rydberg atoms trapped onto a one dimensional optical lattice. In the sector with vanishing total magnetization exact calculations associated to bosonization technique allow to predict that both metallic and superconducting topological Haldane states can be achieved. With the addition of an appropriate magnetic field the system enters a domain wall structure with finite total magnetization. In this regime the conducting Haldane states are confined in domains separated by regions where a fully polarized Luttinger liquid occurs. A procedure to dynamically stabilize such topological phases starting from a confined Ising state is also described.
\end{abstract}

\maketitle

\paragraph{Introduction} Interacting quantum phases with topological features \cite{wen,wen2,wen3,turner,wen4,sentil} associated with a class of non-local excitations represent one of the most intriguing topics in different areas of quantum physics. Besides providing a first case of quantum regimes which go beyond Landau's theory of phase transitions, it has been further demonstrated that these peculiar excitations are deeply connected with symmetry protected topological order \cite{Kit,PTBO}. A well known example is the Haldane phase occurring in spin-1 $XXZ$ chain \cite{Haldane1983} where the presence of fractionalized edge states is captured by a non-local string operator \cite{Nijs1989}. Relevantly a similar behavior is observed in other paradigmatic spin Hamiltonians like AKLT \cite{aklt} and colorless Motzkin chains \cite{barbiero2}.\\
Haldane states with topological features can be induced by interaction also in itinerant bosonic \cite{dallatorre,dallatorre2,batrouni,rossini} and fermionic \cite{montorsi,nonne,barbiero1,MDIR,ota} systems. Due to both the high level of parameter control and the fact that string order parameters can be measured by in-situ imaging \cite{Endal,hilker}, many body ultracold atomic systems \cite{bloch} represent a very promising platform where these models and their topological states can be achieved. Indeed many proposals involving particles with long range dipolar interaction \cite{lahaye} trapped in optical lattices have been presented \cite{dalmonte,deng,fazzini,fazzini1,barbiero3}. Crucially in all these possible setups Haldane orders are associated with a finite value of the charge gap thus describing symmetry protected topological insulators. On the other hand it has been recently shown that a challenging implementation involving polar molecules can support the appearance of Haldane orders with a gapless charge sector \cite{fazzini3}. In this special configuration metallic states as well as regimes with dominant superconducting correlations can be achieved by manipulating the interaction strength between molecules in different dressed rotational states. Beside the fact that a high density sample of polar molecules has been recently achieved \cite{demarco} and spin exchange processes observed \cite{yan}, the experimental implementation of the model supporting Haldane superconductivity remains very challenging.\\
\begin{figure}
\includegraphics[trim={6.5cm 21.5cm 5.cm 1.5cm},clip]{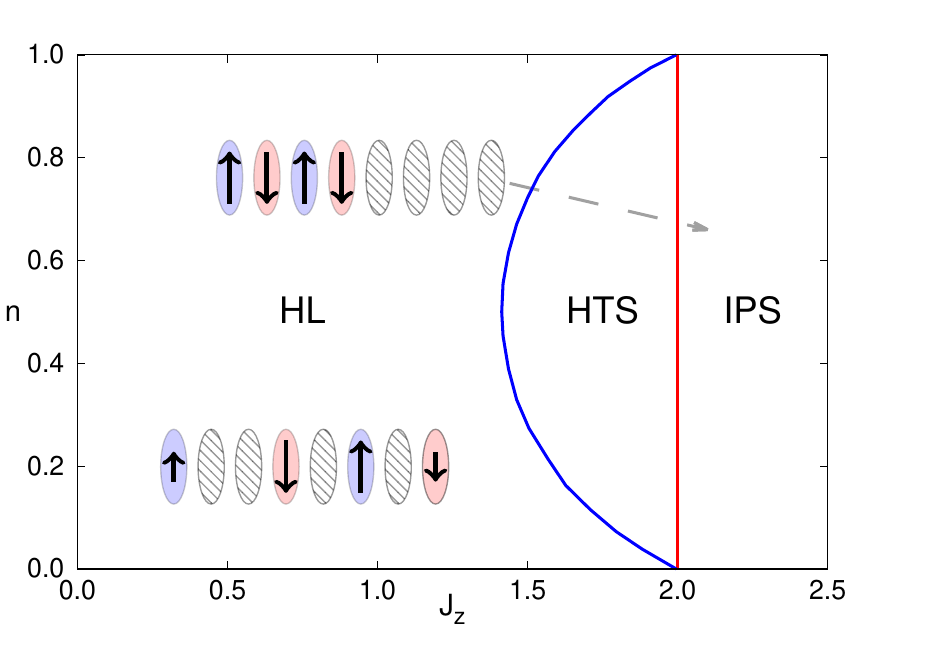}
\caption{	\label{fig:ph_diag_d0} Phase diagram of the $t-J_z$ model eq. (\ref{ham1}) with $\delta=0$ as a function of the filling $n$ and $J_z$. Here the topological conducting phase ($J_z<2$) is denoted as Haldane liquid (HL), hosting in particular a Haldane triplet superconducting regime (HTS, see text). The state with $J_z>2$ is characterized by phase separation into an Ising antiferromagnet and empty sites, and is denoted as IPS. The cartoons reproduce a generic state in the HL and IPS phases; the three possible values of $S_i^z$, namely $+1,0,-1$ are depicted as $\uparrow,0,\downarrow$ respectively. The shorter arrows at the edges emphasize the fractional edge spins, in this case $+n$ at left and $-n$ at right.}
\end{figure}
At the same time Rydberg atoms represent a very promising and flexible platform where long-range interaction can allow the study of intriguing properties of matter. Among many of them it is worth to mention recent results where crystal structures \cite{rydberg1}, $Z_N$ density waves \cite{rydberg2}, interacting topological insulators \cite{rydberg3} and Schr\"odinger cat states \cite{rydberg4} have been experimentally observed. Moreover recent impressive advances in manipulating such atoms allowed the production of ``dressed" Rydberg excitations  \cite{gross1,gross2}. In this setup the life time of the Rydberg excitations is able to exceed the time required by a particle to tunnel between two nearest neighbour sites of an optical lattice. In the case of Ising-like spin-spin interactions, already observed in laboratory, such a aspect allows in principle for the experimental realization of the usually called $t-J_z$ model \cite{ZKMS}, whose topological properties have only recently been studied in the context of a generalized $t-J$ model \cite{fazzini3}.\\
Motivated by this intriguing experimental platform here we address in details the topological properties of the aforementioned $t-J_z$ model. In the case of nearest neighbour couplings  the model is integrable and some ground state properties can be extracted exactly. In particular, as a function of the filling and the Ising coupling, both a spin gapped liquid and a phase separated regime are present. Moreover, in the liquid phase, it is known that, for sufficiently large $J_z$, dominant superconducting correlations can take place \cite{batista}. On the other hand, the bosonization analysis reported in \cite{fazzini3} predicts topological features for such liquid in both the conducting and superconducting regimes: both should be hosted in a topological Haldane ground state, with non-vanishing spin string order correlation function and fractionalized spins at the edges. Thus the integrability of the model allows us to exactly derive the values of these topological quantities. In addition to this we investigate by density-matrix-renormalization-group (DMRG) \cite{white} numerical analysis further quantities to enforce our findings related to the topological aspects of the model. As a first countercheck we capture the constant finite value of the string order parameter as well as a double degenerate entanglement spectrum \cite{Kit,turner,turner2,io2} which are both characteristic signatures of symmetry protected topological states. Then we concentrate on the effect of a finite magnetic field which enriches the complexity of the treated system. Our numerical simulations show that for a weak magnetic field the system remains in the sector of total vanishing magnetization. On the other hand stronger fields produce a finite magnetization, namely an exceeding number of particles in one spin state than in the other. Although the system remains conducting, it forms a domain wall structure composed by distinct regions with different features: a fully polarized Luttinger liquid (FPLL) of particles in just one internal state and a Haldane topological region with vanishing total magnetization. For larger couplings the latter phase is replaced by an insulating antiferromagnetic Ising domain with effective unit density. Finally, by means of time-dependent DMRG \cite{white2,white3} we show how such topological states can be dynamically generated and stabilized by performing an expansion procedure, which is a well established technique in cold atomic experiments, see for instance \cite{bloch2,bloch3} .   

\paragraph{Model}
As shown in \cite{gross1,gross2} dressed Rydberg atoms interact via an Ising antiferromagnetic coupling. Due to the large life time of these excitations tunneling processes at non-unit fillings have to be taken into account. In the case of nearest neighbour coupling which has been shown to be an accurate approximation in very similar models \cite{manmana,fazzini3}, these features are described by the so called $t-J_z$ model

\begin{eqnarray}
H=&-&t\sum_{i,\sigma} \left (c^\dagger_{i,\sigma} c_{i+1,\sigma}+h.c.\right )+
\nonumber\\
&+&\sum_{i}\left(J_z\,S_i^zS_{i+1}^z+\delta \,S_i^z\right)
\label{ham1}
\end{eqnarray}
describing a system of $N=N_\uparrow+N_\downarrow$ particles \cite{note} loaded onto $L$ sites, with total density $n=N/L$. In particular $c^\dagger_{i,\sigma}$ creates a particle in the internal or, analogously, spin state $\sigma$ at the $i$-th site and $S^z_i\doteq(n_{i,\uparrow}-n_{i,\downarrow})$. Besides $t=1$ which fixes our energy and time scale and characterizes the hopping processes of a particle tunneling in a nearest neighbor (NN) site, the other coupling constant $J_z$ describes antiferromagnetic exchange in the $z$ plane, and $\delta$ has the role of a magnetic field which fixes the total magnetization $S^z_{tot}=\sum_i S^z_i$. Beside the fact that the aforementioned parameters are independently tunable, we also impose that double occupancies are strictly forbidden which is a regime easily achievable by using Feshbach resonances. 

Interestingly, the $t-J_z$ Hamiltonian at $\delta=0$ has been introduced decades ago to model the behavior first of heavy fermions compounds and later of high-$T_c$ superconducting materials in the regime of strong interaction. In particular, it allows to emphasize how restricted charge fluctuations induce further spin correlations in these systems (see \cite{ZKMS} and references therein) away from half-filling, where the system is conducting. While spin and charge structure factors in the ground state, as well as correlation functions of other local quantities, have been studied thoroughly, so far very little attention has been given to the topological aspects of the model. Recently these have been unveiled in the context of the numerical investigation of a more general model, namely the $t-J_z-J_\perp$ model \cite{fazzini3}. There the results, also supported by a bosonization and renormalization group analysis, identified in the weak $J_\perp$ regime the presence of Haldane ordering of the spin degrees of freedom. This is captured by a non vanishing value in the thermodynamic limit of solely the spin string order parameter, namely:
\begin{equation}
{\cal O}_S^s(r)\doteq -\langle S_i^ze^{i\pi \sum_{j=1}^{r-1} S_{i+j}^z }S_{i+r}^z\rangle \hspace{2pt}. \label{string}
\end{equation}
A non zero value of the above quantity can be recognized to describe the alternation along the chain of the spins of the particles, diluted in a background of empty sites, as depicted in fig \ref{fig:ph_diag_d0}. 

The ground state energy of the $t-J_z$ Hamiltonian eq. (\ref{ham1}) at $\delta=0$, and some of its properties, have been derived exactly in \cite{batista}. There it was noticed that, since doubly occupied sites are forbidden, singly occupied sites are strictly antiferromagnetically ordered at any $J_z>0$. In this way the problem can be mapped into that of finding the ground state of an equivalent attractive spinless fermions model. More precisely, the mapping amounts to replace in the kinetic term $c_{i\sigma}\rightarrow f_i$, $f_i$ being a spinless fermion annihilation operator. As for the Ising-like interaction term, one realizes that it is equivalent to an attractive density-density term: $S_i^z S_{i+1}^z\rightarrow - n_i^f n_{i+1}^f$, with $n_j^f=f_j^\dagger f_j$ . The resulting attractive $t-V$ model ($V=-J_z$) reads: 

\begin{equation}
H=-t\sum_{i} (f^\dagger_{i} f_{i+1}+h.c.)+V \sum_{i}n_i^fn_{i+1}^f
\label{tV}
\end{equation}
whose $T=0$ phase diagram is well known. For $V<0$ ($J_z>0$) it is characterized by two critical values: when $J_z>2$, a phase separated state occurs, where the singly occupied sites separates from the empty ones. Whereas for $J_z<2$  the ground state is a liquid, with dominant superconducting correlations at large distance when the Luttinger exponent $K_c>1$, i.e.  for $J_z^{c}<J_z<2$. In other words, this regime is characterized by the fact that the triplet superconducting correlation function 
\begin{equation}
\label{triplet}
O_{TS}(r)=\langle C_{TS}^\dagger(i)C_{TS}(i+r) \rangle
\end{equation}
where $C_{TS}^\dagger(i)=\frac{1}{\sqrt{2}}(c^\dagger_{i,\uparrow}c^\dagger_{i+1,\downarrow}+c^\dagger_{i,\downarrow}c^\dagger_{i+1,\uparrow})$, goes to zero with $r$ slower with respect to the other correlation functions having a power law decay. Relevantly both $K_c$ and $J_z^c$ can be computed exactly by Bethe ansatz \cite{Hal,QFYal}. In particular, at filling $n=1/2$, one gets
\begin{equation}
K_c=\frac{\pi}{4(\pi-\mu)}
\end{equation}
with $\cos{\mu}=-\frac{J_z}{2t}$. 

\paragraph{Ground state topological properties at $\delta=0$} Exploiting the rigorous results of the previous section on the phase diagram of the $t-J_z$ model at $\delta=0$, we can derive exactly the non-vanishing value of the spin string parameter \eqref{string} and other topological properties. In fact, restoring in the ground state of the above $t-V$ model eq. (\ref{tV})the alternation of spin orientation of the fermions, one realizes that the phases must be also spin gapped. In agreement with the general bosonization analysis reported in previous work \cite{fazzini3}, the spin gap should be characterized by a non-vanishing value of the Haldane spin string order eq. (\ref{string}).
\\Since in the ground state the spins of the occupied sites are strictly alternating, the string at the exponent in \eqref{string} is either $0$, in case where the two spins at the edges are opposite, or $\pm1$, in case they are parallel. Thus we can rewrite the Haldane string order parameter as:
\begin{equation}
{\cal O}_S^s(r)=\langle (-)^{\frac{1}{2}(S_i^z-S_{i+r}^z)}S_i^zS_{i+r}^z\rangle \quad . 
\end{equation}
The right hand side here is such that contributions to the expectation value coming from opposite spins at the edges are identical to those coming from parallel spins. Thus  ${\cal O}_S^s(r) \equiv\langle |S_i^z| |S_{i+r}^z|\rangle$. The quantity can now be expressed in terms of the density-density correlation function\cite{LuPe} of the spinless $t-V$ model \eqref{tV}. Explicitly, since the average density $n$ of the spinless fermions is the same as that of the spinfull liquid, in the thermodynamic limit we have:

\begin{equation}
{\cal O}_S^s(r)\doteq \lim_{r\rightarrow\infty} \langle n_i^f n_{i+r}^f\rangle = n^2\quad .\label{string_inf}
\end{equation}

On general grounds, in case of Haldane spin string order, for an open chain of length $L$ one expects two degenerate distinct ground states, $|\psi\rangle_\pm$. They are characterized by opposite fractional spins accumulated at the two edges \cite{MDIR}, as schematically shown in the cartoon of Fig. \ref{fig:ph_diag_d0}. Its value for the Haldane ground state of the $t-J_z$ model again can be calculated using the correspondence with the $t-V$ model.  Indeed at zero total magnetization, for each of the two degenerate topological states, the probability of having a particle with spin $\pm 1$ in one of the system edges ($i=1$ or $L$) is equal to the probability of having a spinless fermion, so that

\begin{equation}
\langle S_1^z\rangle_\pm =\pm  \langle n_1^f\rangle= \pm n=-\langle S_L^z\rangle_\pm \quad , \label{edge} 
\end{equation}
which is in fact non vanishing and smaller than one at any filling $n\neq 1$ and thus fractionalized. Also, manifestly, the value of each of the two edge spins is different from the bulk magnetization and correlated to the other.

The aforementioned results are summarized in Fig. \ref{fig:ph_diag_d0}. In particular the latter shows a new and more complete description of the phase diagram reported in \cite{batista} where the topological properties of eq. (\ref{ham1}), as already specified, were unknown.
\\More precisely, away from unit density we have three possible different phases. For $J_z<2$ we observe a phase with no charge gap and a finite value in the thermodynamic limit of the spin string parameter given by \eqref{string_inf} thus being a liquid phase with topological features that we denote as Haldane liquid (HL) phase. In fact, the HL phase hosts two regimes: for $K_c<1$ (i.e. $J_z < J_z^c$) it has dominant spin-spin correlations
\begin{equation}
C_s(r)=\langle S^z_iS^z_{i+r}\rangle-\langle S^z_i\rangle\langle S^z_{i+r}\rangle \hspace{2pt}.
\end{equation} 
Whereas for $K_c>1$ (i.e. $J_z^c<J_z<2$) it has dominant triplet pair-pair correlations (\ref{triplet}). To emphasize its superconducting properties, we denote this latter regime of the HL phase as Haldane triplet superconductor (HTS). Finally, for $J_z>2$, where the spinless $t-V$ model eq. (\ref{tV}) enters the phase separated state, correspondingly in the $t-J_z$ model particles get separated from empty sites and a phase separated antiferromagnetic Ising phase (IPS) takes place. 

It is worth mentioning that the above HL phase is an example of a larger class of one dimensional conducting phases of fermions characterized by non local order in the spin channel. The most renown is the Luther Emery phase, entered when spin parity order becomes finite\cite{barbiero1,MDIR}. It was recently observed that parity orders can be generalized to two dimensional arrays \cite{FBM,TBM}, in particular capturing the presence of a phase qualitatively similar to the Luther Emery liquid also in this case. On the contrary, it is expected that the topological nature the HL phase severely limits the possible generalization of string orders to higher dimension\cite{AnRo}.

\paragraph{DMRG analysis.}
\begin{figure}
\includegraphics[scale=0.35]{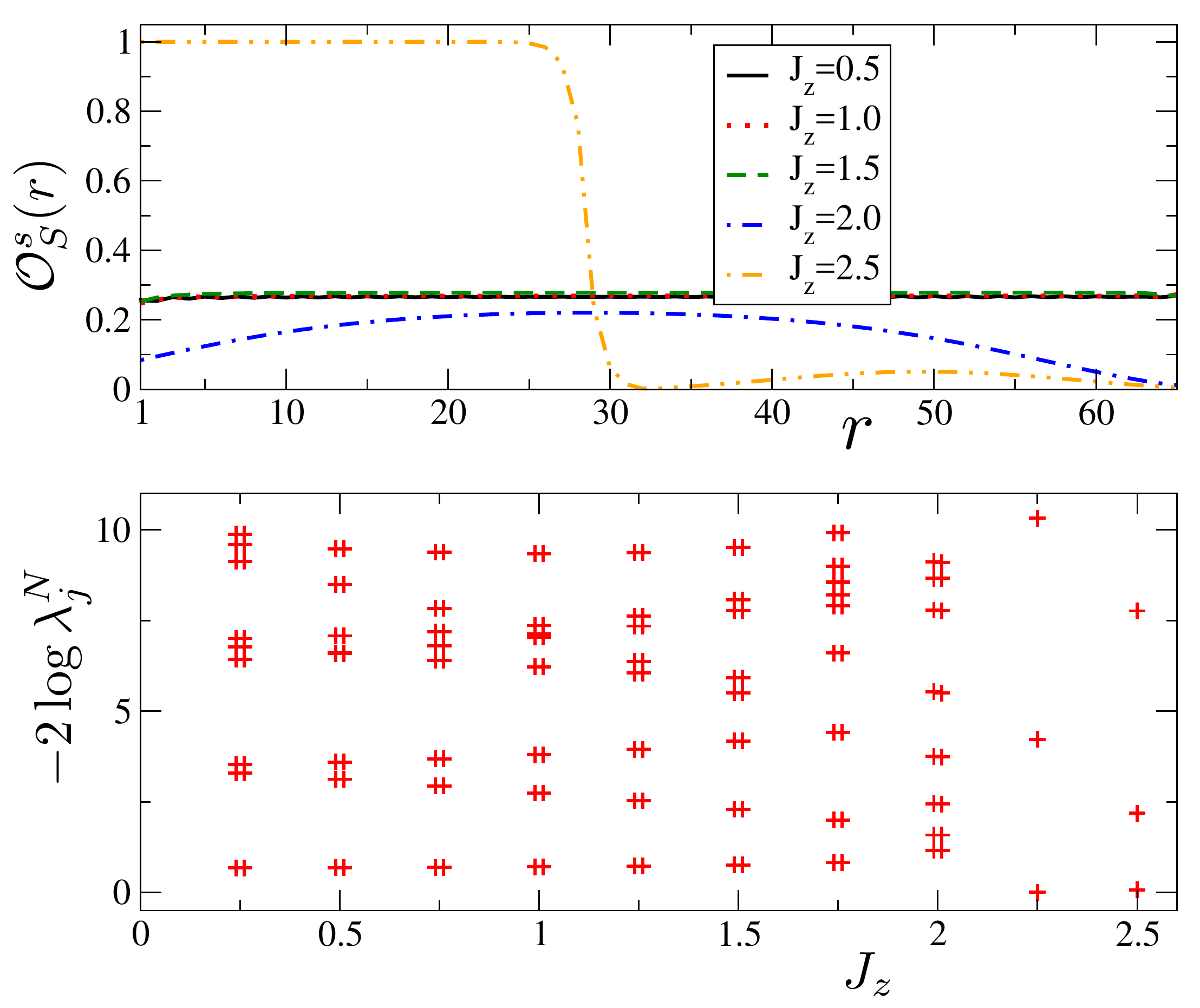}
\caption{\textit{Upper panel} decay of ${\cal{O}}_S^s(r)$ for different $J_z$ and $\delta=0$ with $L=81$, $N=(L+1)/2$ particles and keeping only the central sites of the system. In order to pin the Ising domain in the right part of the lattice we also include weak antiparallel magnetic fields at the systems edges.  \textit{Lower panel} values of the entanglement spectrum for different $J_z$ and $\delta=0$ with $L=81$ and $N=(L+1)/2$ particles.}
\label{es}
\end{figure}
In order to support 
the analytical results of the previous paragraph, here we employ DMRG technique. Except when explicitly mentioned, in all the performed simulations we use up to 768 dmrg states and perform up to 6 finite size sweeps which ensure convergence with a truncation error smaller than $10^{-9}$. 
As a first step we evaluate the decay of the string correlation eq. (\ref{string}) when moving from the topological phase to the phase separated regime. As visible in Fig. \ref{es} we find that this quantity is constant in the phase where we predicted the presence of fractionalized edge states and its value is already quite close to the one expected in the thermodynamic limit  \eqref{string_inf}. On the other hand at the transition point $J_z=2$ this peculiar behavior is not present anymore. Moreover once the system enters the phase separated state the string approaches the value ${\cal O}_S^s=1$ in the region where true Ising antiferromagnetic order occurs while it goes to zero in the region where only empty sites are present. 
\\A further probe signaling Haldane topological orders is the entanglement spectrum which corresponds to the eigenvalues $\lambda^N_j$ of the reduced density matrix $\rho_A=\sum_{Nj}\lambda^N_j\rho_j^N$ with respect to some system bipartition $A$, where $\rho^N$ describes a pure state of $N$ particles. In particular it is known that in a topological phase the entanglement spectrum has to show even degeneracy in the lower $\lambda_j^N$s \cite{Kit,turner,turner2,io2}. Fig. \ref{es} indeed confirms that in the phase where edge states are predicted we get a perfect two-fold degeneracy. At the transition point the degeneracy is not accurate anymore and it totally disappears in IPS. 

\paragraph{Effect of finite magnetization.}

\begin{figure}
\includegraphics[scale=0.37]{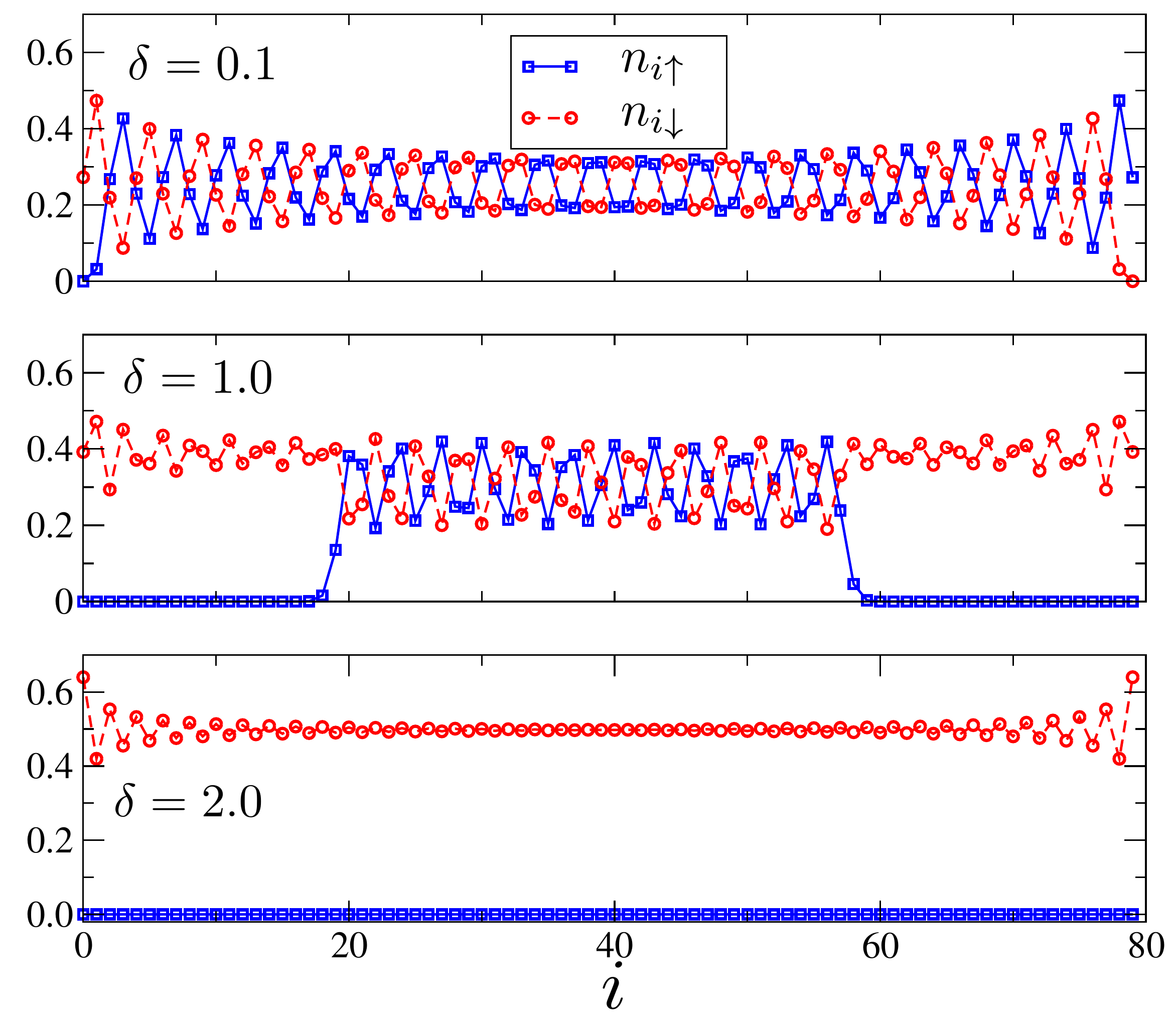}
\caption{Density distribution for different values of $\delta$ at fixed $J_z=1.5$ $L=80$ and $N=L/2$. In all cases a weak positive local magnetic field $\mu$=0.01 is applied in the system center to lift the GS degeneracy.}
\label{findelta}
\end{figure}

Finite values of $\delta$ strongly enlarge the richness of the aforementioned phase diagram (Fig. \ref{fig:ph_diag_d0}). Intuitively the role of a positive magnetic field is to make energetically more favorable for the system to have more particles in one internal state than in the other, say $N_\downarrow>N_\uparrow$ with $N_\sigma=\sum_i n_{i\sigma}$. Of course this contrasts with the effect of $J_z$ which for $N_{\uparrow}=N_{\downarrow}$ can maximize the number of antiferromagnetic surfaces. As shown in Fig. \ref{findelta} our results indicate that for weak $\delta$s the system still prefers the solution with vanishing total magnetization, thus supporting the presence of the same phases as in Fig \ref{fig:ph_diag_d0}, where for $J_z<2$ the Haldane topological orders extend all over the lattice sites and for $J_z>2$ IPS takes place. 
An unbalanced phase with $S^z_{tot}\neq 0$ occurs only above a critical value of the magnetic field, almost independent of the strength of $J_z$. 
In this regime, as the example reported in the central panel of Fig. \ref{findelta} shows, the system prefers to arrange into a domain wall structure composed by two regions: one with $S^z_{tot}=0$ and the other being a fully polarized Luttinger Liquid. More precisely we get that, for $J_z<2$, the whole system remains conducting hosting a uniform finite amount of holes, though spatially distinguished regions with different magnetization can be noticed: one with  $N_{\uparrow}=N_{\downarrow}$ and the other fully polarized ($N_\uparrow=0$). 
\begin{figure}
\includegraphics[scale=0.36]{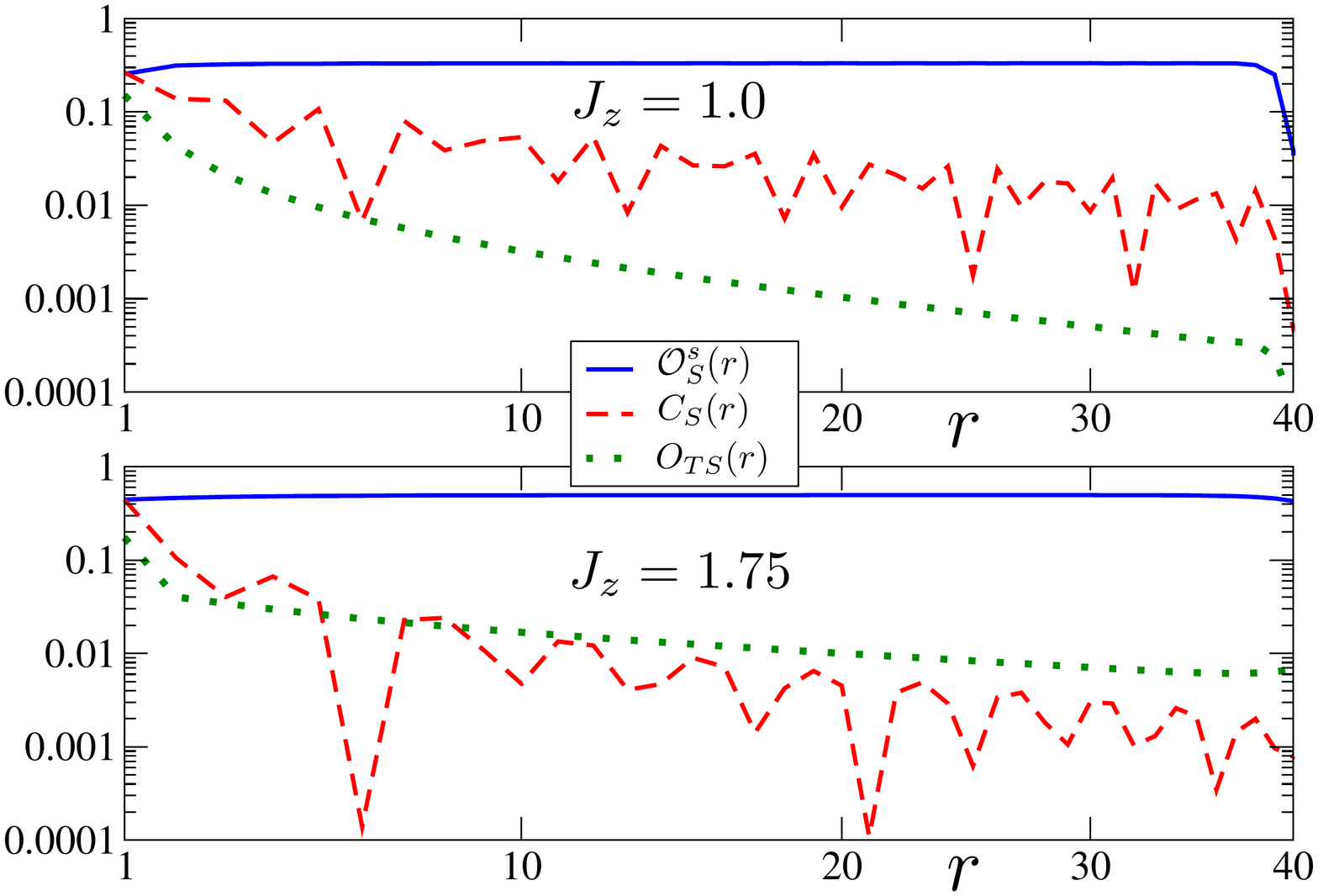}
\caption{Decay of ${\cal{O}_S^s}(r)$, $O_{TS}(r)$ and $C_S(r)$ for $\delta=0.4$ $L=80$ $N=L/2$ for two different values of $J_z$. The correlations are evaluated in the central region of the system where $N_\uparrow=N_\downarrow$. In both cases a weak positive local magnetic field $\mu$=0.01 is applied in the system center to lift the GS degeneracy.}
\label{figsc}
\end{figure}

Both coexisting states are the ground state of an effective Hamiltonian of the form of the spinless $t-V$ Hamiltonian eq. (\ref{tV}) with one crucial difference. For the fully polarized state, besides an additional constant term, the effective interaction is now repulsive ($V=J_z>0$). Thus the region amounts to a fully polarized Luttinger liquid with $K_c<1$. The unpolarized region instead preserves its topological features since the effective model capturing such a domain is exactly eq. (\ref{ham1}). At $n=0.5$, both states have the same filling of the uniform phase: the spatial extent of each of the two is determined by the strength of $\delta$. The central panel in Fig. \ref{findelta} indeed shows the presence of a central region with filling lower than one and diluted antiferromagnetic order; for the parameters value reported in figure it turns out to be a HTS region (HL in case of lower $J_z$). Whereas in the left and right parts two FPLLs are seen. A further increase in $\delta$ makes the presence of $\uparrow$ particles too energetically costly and a fully polarized liquid with $N_{\downarrow}=N$ and $N_{\uparrow}=0$ occurs in the whole lattice. In order to prove that the lattice part with vanishing magnetization is described by eq. \ref{ham1}, we show in Fig. \ref{figsc} the behavior of the correlation functions evaluated in such a region. In particular on one hand it is possible to notice that ${\cal{O}}_S^s(r)$ is always constant thus confirming the topological nature of the domain with $N_\uparrow=N_\downarrow$. On the other hand we find that the triplet superconducting correlation function becomes the leading order only when $J_z^c<J_z<2$ thus confirming the validity of our analysis.\\ The situation becomes different when a Ising coupling $J_z>2$ is considered. Indeed, as known from the study of the $\delta=0$ case, such a strong interaction destroys the Haldane order in favor of a true Ising antiferromagnet where particles and holes are totally demixed. In this context the magnetic field plays a similar role as in the previous case. In particular for weak $\delta$s the system remains balanced, whereas above a critical value $\delta_c$ a domain wall structure composed by an Ising antiferromagnet and a FPLL  is obtained. Moreover, as in the already discussed scenario, large magnetic fields drive the system in a regime where only $\downarrow$ particles are present and due to the specific choice of the considered density $n=0.5$ a gapped phase with density wave (DW) order takes place. It's relevant to underline that for any other filling DW is substituted by a LL as in the previous case.  

\begin{figure}
\includegraphics[scale=0.7]{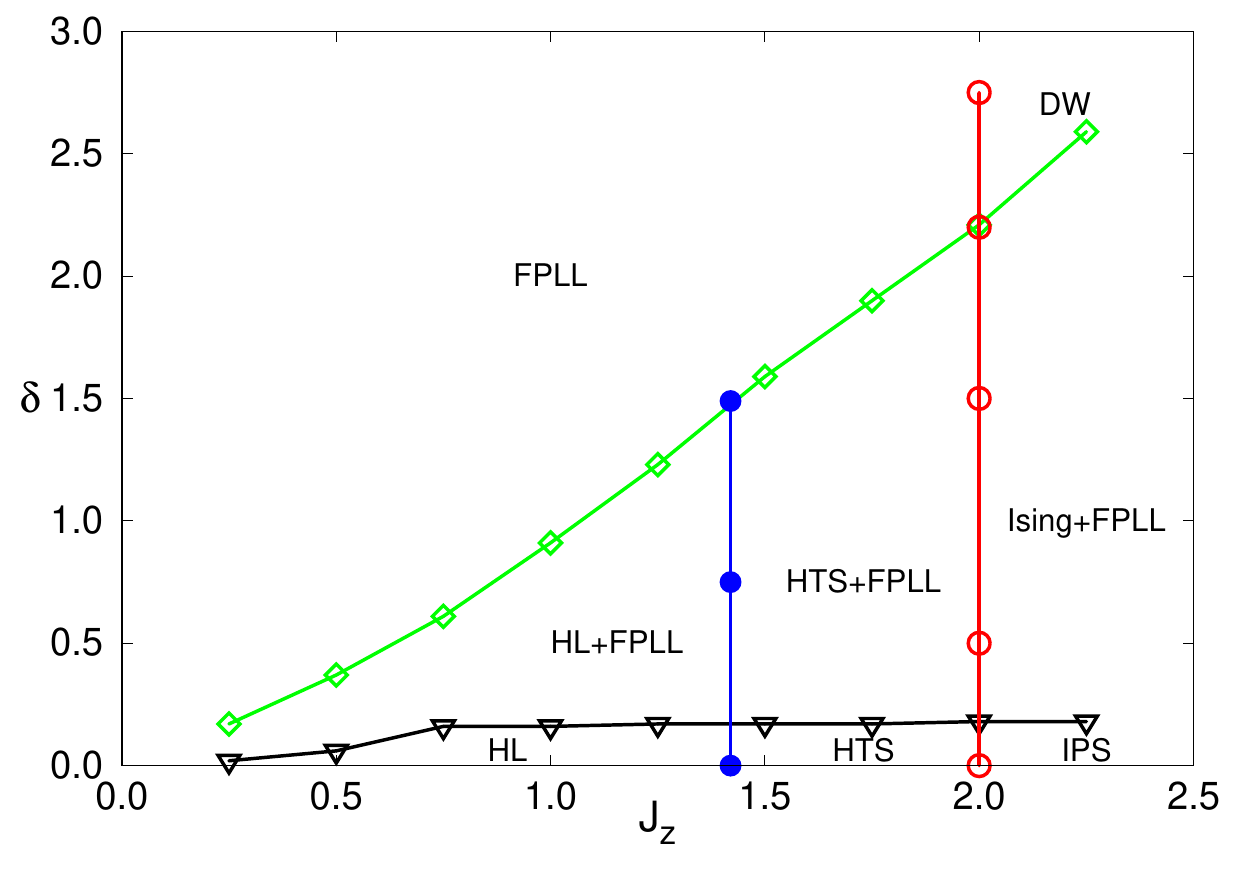}
\caption{Phase diagram of the $t-J_z$ model eq. eq. (\ref{ham1}) as a function of $\delta$ and $J_z$ with density $n=0.5$ and $L=80$. Here also regions characterized by phase separated states are reported: the $+$ symbol is used to emphasize this aspect.}
\label{pdd}
\end{figure}

The above discussed behavior is summarized in the phase diagram shown in Fig. \ref{pdd}. Here it is possible to notice that a weak magnetic field does not alter the phases of Fig. \ref{fig:ph_diag_d0} till a critical value of $\delta$ where a regime of phase separation with finite total magnetization takes place. Here, beside a fully polarized region of particles behaving as a LL, as function of the strength of $J_z$ three possible phases can appear, a Haldane liquid (HL), a Haldane phase with dominant superconducting order (HTS) and an Ising antiferromagnet (Ising). On the other hand larger values of $J_z$ and $\delta$ give rise, due to the specific considered density, to an insulating DW state. We observe that the slope of the transition line to either FPLL or DW (green line) increases with $J_z$ and, in fact, it is expected to approach the value $2$ in the strong coupling limit. Moreover we underline that for higher total fillings and $J_z<2$ domain wall structures composed by both Haldane orders and by DW domains can happen.

\paragraph{Dynamical stabilization of the Haldane states.}
The above Haldane conductors can be dynamically stabilized in experiments involving dressed Rydberg states. Our starting point is precisely the state that has been already reached in such experiments, namely an Ising antiferromagnet with exactly one particle per lattice site. Due to the infinite on-site repulsion between particles in such an Ising state, tunneling processes are not energetically possible. In order to make Hamiltonian eq. (\ref{ham1}) a correct description of such a setup, holes, having the role of restoring the hopping processes,  have to be injected in the system. A way to achieve that is to use optical tweezers \cite{schlosser} which can selectively remove one by one atoms in the lattice and hence generate empty sites. Another way to make the Rydberg excitations move along the lattice is to simply let the Ising state expand. In particular our employed strategy is to prepare an initial state where all the atoms are compressed in the central part of the lattice by a strong harmonic confinement which, in this region, excludes the presence of empty sites, like for instance $|00..\uparrow\downarrow\uparrow\downarrow..00\rangle$. 
\begin{figure}
\includegraphics[scale=0.35]{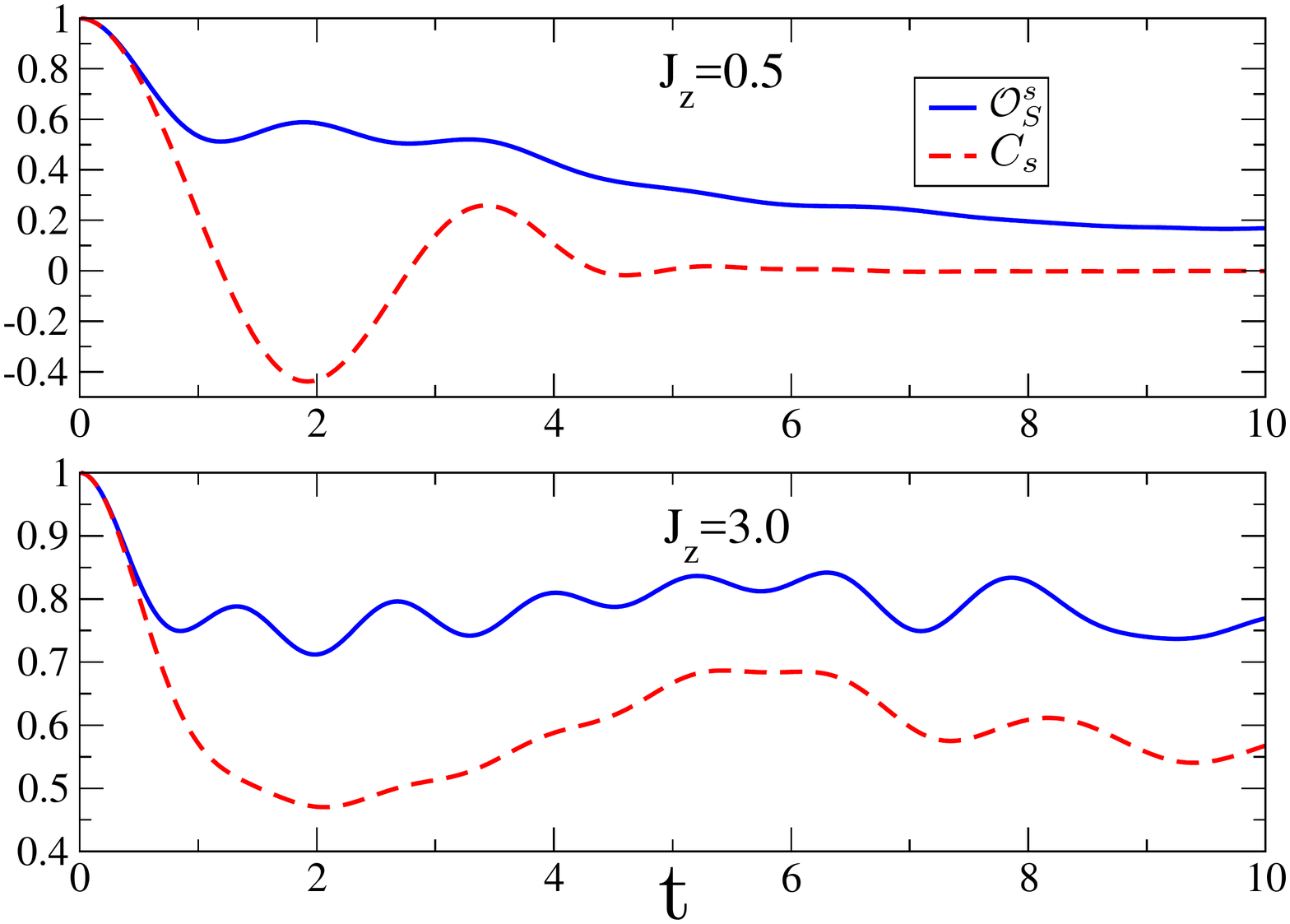}
\caption{\textit{Upper panel} Time evolution of $C_s$ and ${\cal{O}}_S^s$ in a system with $L=40$ and $N=L/2+1$ particles and starting with and initial state with fixed $J_z$ and a strong harmonic potential to confine the particles in the central part of the system. The evolution is performed by keeping the same value of $J_z$ and removing the harmonic potential. Both quantities are evaluated in the central 10 sites of the systems. The numerical simulation are performed by keeping up to 512 dmrg states in the evolution and a time step $\delta \text{t}=0.01$}
\label{dyn}
\end{figure}

Due to the finite value of $J_z$ a perfect Ising antiferromagnet can be reached in the central part of the lattice. Once this state has been prepared we remove the harmonic confinement. This has the effect of letting the atomic cloud expand and thus changing the effective particle density in the lattice. In this way the system will equilibrate in a state whose physics will be captured by eq. (\ref{ham1}). In order to understand the validity of such an approach we monitor the evolution in time $\text{t}$ of the string order parameter eq. (\ref{string}), and of the spin-spin correlation function $C_s$. The latter should remain finite only when true Ising antiferromagnetism is present. Fig. \ref{dyn}, where for simplicity we kept $\delta=0$, shows indeed that such a procedure is able to stabilize Haldane conductors. In particular in the upper panel the value $J_z=0.5$ supports the presence of a topological state and clearly our results demonstrate that the string remains finite during the whole evolution and stabilizes to a non vanishing value. On the other hand $C_s$ displays oscillations and after a certain time reaches a constant vanishing value. This last feature changes when larger $J_z$ are considered. Indeed the lower panel of Fig. \ref{dyn} shows the time dependent expectation values of both the considered quantities as before but for an Ising coupling $J_z=3$ which supports the presence of a true antiferromagnet. Here clearly both $C_s$ and ${\cal O}_S^s$ remain finite during the evolution confirming the fact that the system will thermalize in a phase separated state with an Ising domain in its central part (IPS).  

\paragraph{Conclusions}

We have shown that Rydberg dressed atoms trapped onto a one dimensional optical lattice represent an ideal candidate to probe Haldane topological orders with conducting features. In particular, for a vanishing magnetic field, the phase diagram can be derived analytically and the topological properties are extracted in an exact way. At the same time our numerical results allow to both confirm the analytically predicted topological nature of the studied model and to investigate other peculiar cases. When a finite magnetic field is applied, we observed that the system arranges in a domain wall structure.  For not too large $J_z$ the domains turn out to be composed by a fully polarized Luttinger Liquid and by a Haldane topological state whose conducting or superconducting properties are determined by the strength of the antiferromagnetic coupling. Whereas at larger $J_z$ the Haldane state is replaced by an Ising domain with true antiferromagnetic order. Moreover strong enough magnetic fields destroy the domain structure and, depending on the antiferromagnetic coupling, either homogeneous Luttinger Liquids or, in specific cases, a density wave state occur. We also demonstrate that a protocol based on atomic expansion can be experimentally employed to dynamically stabilize such Haldane conductors. The present results enrich our understanding on interaction induced symmetry protected topological states of matter, giving a reliable procedure to experimentally reach and study such features with the currently available experimental platforms. 

\begin{acknowledgments}
{\it Acknowledgments:} 
The authors thank I. Bloch, C. Gross, F. Grusdt, and T. Macr\'i for interesting discussions. L. B. acknowledges ERC Starting Grant TopoCold for financial support. After completion of the current manuscript we got aware of a recent preprint where topological aspects of the $t-J_z$ model are analyzed \cite{sala}.

\end{acknowledgments}


\begin{thebibliography}{}
\bibitem{note} Here we consider particles with fermionic statistic but exactly the same results are expected in the case of a two component Bose gase with hard-core onsite repulsion.
\end{thebibliography}
\end{document}